\documentclass[11pt]{article}

\usepackage{geometry}
\usepackage{currfile}
\usepackage{textcomp}

\usepackage{xcolor,soul}
\sethlcolor{yellow}

\usepackage[pdftex]{graphicx}
\DeclareGraphicsExtensions{.pdf,.jpeg,.png}
\graphicspath{{./.}}

\date{May 31, 2019}

\begin{document}


\title{ \LARGE{A Federated Authorization Framework\\
for Distributed Personal Data \& Digital Identity}\\
~~}
\author{
\large{Thomas Hardjono}\\
\large{~~}\\
\large{MIT Connection Science \& Engineering}\\
\large{77 Massachusetts Avenue}\\
\large{Cambridge, MA 02139, USA}\\
\large{ {\tt hardjono@mit.edu} }\\
\large{~~}\\
}

\maketitle

\begin{abstract}
The digital identity problem is a complex one in large part because
it involves personal data,
the algorithms which compute reputations on the data
and the management of the identifiers that are linked to personal data.
The reality of today is that personal data of an individual
is distributed throughout the Internet,
in both private and public institutions,
and increasingly also on the user's devices.
In order to empower individuals to have a say in who has access 
to their personal data and to enable individuals to 
make use of their data for their own purposes,
a coherent and scalable access authorization architecture
is required.
Such an architecture must allow different data holders, data providers and user-content generators
to respond to an individual's wishes with regards to consent
in a federated fashion.
This federation must allow an individual to easily manage access policies
and provide consent as required by current and forthcoming data privacy regulations.
This paper describes the User Managed Access (UMA) architecture and protocols
that provide the foundation for scalable access authorization.
~~\\
\end{abstract}

\newpage
\clearpage

\section{Introduction}

The advent of blockchain technology has again
brought to the forefront the question of digital identity. 
In order to transact securely on a blockchain system and have 
a meaningful engagement on the Internet generally, {\em insights} 
are needed for individuals and parties involved as the basis of trust establishment and risk management.
These insights are typically conveyed as digitally-signed assertions or claims.
However, in order for algorithms to compute and derive meaningful insights, 
data is needed for these algorithms.
As such, one key dimension of the digital identity problem 
is the challenge of providing privacy-preserving
scalable access to data distributed (siloed) throughout the Internet.

Another dimension of the digital identity problem is the privacy of personal data.
Over the last decade there has been a continuing decline in trust on the part of individuals
with regards to the handling and fair use of personal data~\cite{WEF2011}.
Pew Research reported that 91 percent
of Americans agree or strongly agree that consumers 
have lost control over how personal data is collected and used, 
while 80 percent who use social networking sites are 
concerned about third parties accessing their shared data~\cite{Madden2014}. 
The Webbmedia Group, writing in the Harvard Business Review, 
has identified data privacy as one of the top ten technology trends of 2015~\cite{Webb2015,Maler2015a}.
This situation has also been compounded by the various recent reports of attacks and theft of data
(e.g. Anthem~\cite{Anthem2015}, Equifax~\cite{Equifax2017}).
Related to the loss of trust -- and perhaps as a consequence of it --
is the recent emergence of new regulations aimed at addressing data privacy.
The enactment of the EU GDPR~\cite{GDPR}
has had influenced the data privacy discourse in the United States and elsewhere~\cite{Kerry2019a}.
At the state level, California has issued the California Consumer Privacy Act (CCPA)~\cite{CCPA2018}.

The state of declining trust was already reported
by the World Economic Forum in 2014~\cite{WEF2014}.
The WEF report was the culmination of a multi-year initiative
with global insights from various high level leaders from different
sectors of society (industry, governments, civil society and academia).
A theme running through the 2014 WEF report
is the {\em need to strengthen individual trust}.
The WEF report suggests three means to address this problem ({p.4} of \cite{WEF2014}).
Firstly, increase in {\em transparency} by focusing on engagement and response,
and by providing individuals with insight and meaningful control.
This is instead of the current approaches
focusing on disclosure and
providing details (which often overwhelm individuals).
Secondly, improve {\em accountability} by orienting throughout the value
chain (front-end to back-end) with
risks being equitably distributed.
This is in contrast to the current industry practices that are
oriented towards the front-end
of the value chain with risks
and responsibilities residing
with the individual.
Thirdly, {\em empowerment of individuals} by way giving them a say in how
data about them is used by organizations and 
by giving individuals the capacity
to use data for their own purposes.
Empowerment should be
distributed with shared incentives
for empowering individuals and
distributing value closer to the source
of data production (the individual).
This is in contrast to the current approaches
which are focused on maintaining
information differentials among
a concentrated set of actors.

The recent 2018 Report on Global Consumer Trust from the
Mobile Ecosystem Forum (MEF)~\cite{MEF2018}
points to a number interesting developments with regards to consumer trust in the context mobile devices and apps.
A majority of respondents on this report
have taken one or more actions within the last
two years to protect their privacy and mitigate potential harm.
The report states that 57 percent of users indicate that personal data collection is a risk to them,
and
that 68 percent think it is important to know how their personal data is being used.
The report finds that the personal data market has matured, where smartphone users are ready to do more, including managing
their own data. 
However, tools need to be easier to use and well understood.
For the mobile space, the report finds that
trust is still a key ingredient for the acceptance of mobile services.
A trusted service is more likely to avoid
punitive actions from the user,
such as not downloading an app due to excessive permissions.
As such, there is a correlation between an individual's trust in an organization,
their confidence to control their personal data, and
their use of mobile devices and services.
Empowering people with the ability to control
their identity and personal information is key to earning trust
and customer loyalty.

\begin{figure}[!t]
\centering
\includegraphics[width=0.8\textwidth, trim={0.0cm 0.0cm 0.0cm 0.0cm}, clip]{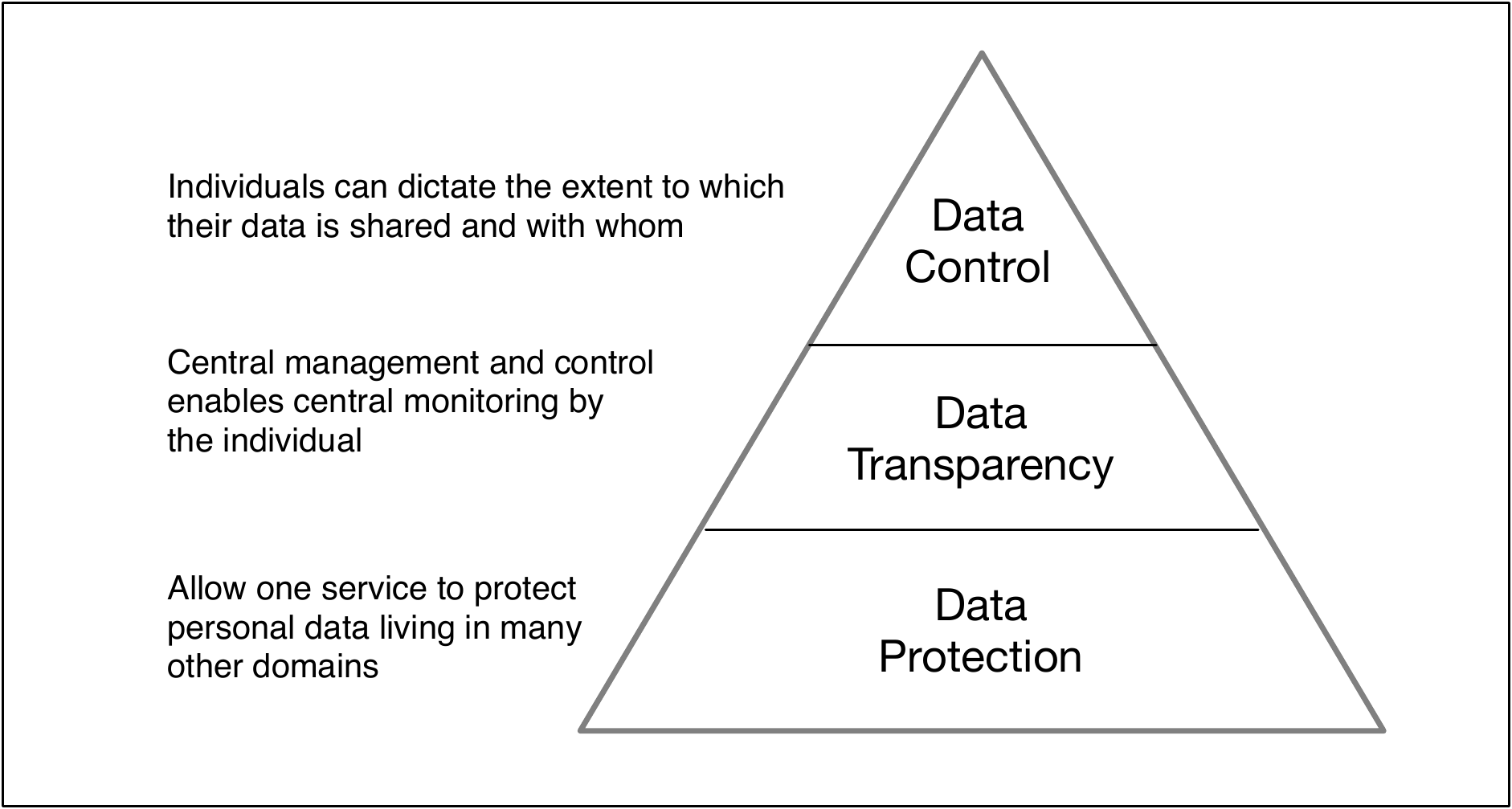}
	%
	%
\caption{The Vision of Privacy {2.0} (after~\cite{Maler2019b}) }
\label{fig:privacy2.0}
\end{figure}

Echoing the WEF Report~\cite{WEF2014} and the MEF findings~\cite{MEF2018},
we believe that individuals need {\em meaningful control} over their personal data,
which is increasingly distributed across various entities on the Internet.
The ordinary user finds the amount of data and complexity of managing data to be overwhelming.
Individuals want transparency in order to understand what data about them is being collected
and what it is being use for.
However, individuals also seek the following (see Figure~\ref{fig:privacy2.0}):
(i) choice in sharing with other parties,
(ii) convenience in sharing/approval with no outside influence,
(iii) centralizable monitoring and management for the individual,
and 
(iv) control of who/what/how at a fine grain~\cite{Maler2019b}.

Today the reality is that personal data -- which are co-created with the individual --
typically does not reside with the individual, for practical reasons.
The GDPR recognized this reality, and reflected it by using the notion of
{\em data controllers} and {\em data processors}\footnote{
A data controller is the natural or legal person, public authority, 
agency or other body which, alone or jointly with others, 
determines the purposes and means of the processing of personal data.
A data processor is a natural or legal person, 
public authority, agency or other body which processes personal 
data on behalf of the controller~\cite{GDPR}.}.
Therefore, in order to provide an individual with true meaningful control over their personal data
the controllers (of an individual's personal data) must collectively
provide an easy way for the individual to configure access policies (consent rules)
that will apply to the personal data located at each of the controllers.
We refer to this as {\em authorization federation}.
The overall goal of authorization federation is to empower the individual
to set access policies (consent permission) at one location (e.g. at one data controller)
and have the access policies propagated automatically to other data controllers
and be enforced there also.
In this way, the individual is relieved from having to log-in to
many sites for the purpose or configuring the access policies multiple times.
This is consistent with and follows from the WEF recommendation
of giving individuals the capacity
to use data for their own purposes.

In the case where a third-party seeks information about an individual,
then such access must be privacy-preserving.
Instead of allowing external third parties to simply read (copy) personal data,
privacy-preserving algorithmic approaches --
such as the MIT Open Algorithms approach~\cite{HardjonoPentland2019b,HardjonoPentland2018a} --
should be deployed.
This approach advocates that (a) data never leaves its repository, 
and that (b) instead the algorithms are transmitted to the data location and be executed there,
and that (c) only vetted algorithms are permitted to be executed.
Thus, data controllers and processors are able to increase 
individual privacy and comply to privacy regulations
by obtaining consent for the execution of algorithms over the personal data
(without moving the data from the controller's respository).
The computed insights can then
be packaged as digitally-signed assertions or claims~\cite{SAMLcore,Sporny2019,W3C-DID-2018}.

This paper focuses on the notion of a {\em federated authorization} model
as applied to personal data held by various data controllers.
We discuss a specific solution called User Managed Access (UMA)
which provides the foundation for the federation among the controllers.

In the next section we briefly review policy-based authorization,
which has been a topic of interest within the computer industry
for over four decades now.
Section~\ref{sec:FederatedAuthNAuthz}
discusses the notion of mediated authentication,
which is today commonly manifested in the form of identity provider services in the consumer space.
We expand the notion of mediation to that authorization
in Section~\ref{sec:DecentFederatedAuthorization},
and describe the UMA authorization architecture
in Section~\ref{sec:UMAarchitecture}.
The paper closes in Section~\ref{sec:Conclusions} with some conclusions.

\section{Policy-Based Access Control and Authorization}
\label{sec:Policy-BasedAccessControl}

The issue of controlling access to multi-user resources has been
an important theme since the mid-1960s,
with the rise of the time-share mainframe computers.
There is today a considerable body of literature in this area,
which is a core part of computer and network security.
Generally, the term {\em access control} is applied not only
to physical access (to the computer systems) but also
to system-resources (e.g. memory, disk, files, etc).
Notable among the early efforts in the late early 1970s
was the Multics system~\cite{Saltzer1974}.
It was Multics which, among others, introduced 
the notion of {\em protection rings} -- a concept 
which has been inherited by many operating systems today.

In the context of government and military applications,
there was the further issue of access based on a person's rank or security clearance.
Here, the the concept of {\em mandatory} and {\em discretionary} access control in multi-level systems
came to the forefront in the form of the Bell and LaPadula Model (BLM)~\cite{BellLaPdula1973}.
In this model, access control is defined in terms of {\em subjects} 
possessing different {\em security levels}, seeking access to {\em objects} (i.e. system resources).
Thus, for example,
in the BLM model a subject (e.g. user) is permitted to access an object (e.g. file)
if the subject's security level (e.g. ``Top Secret'') is higher than security level of the object (e.g. ``Secret'').
The notion of {\em roles} or capacities was added to this model,
leading to the {\em Role-Based Access Control} (RBAC) model.
Here, as a further refinement of the BLM model,
a subject (user) may have multiple {\em roles} or capacities within a given organization.
Thus, when the subject is seeking access to an object,
he or she must indicate the role within which the request is being made.
The formal model for RBAC was defined by NIST in 1992~\cite{FerraioloKuhn1992}.

\begin{figure}[!t]
\centering
\includegraphics[width=0.8\textwidth, trim={0.0cm 0.0cm 0.0cm 0.0cm}, clip]{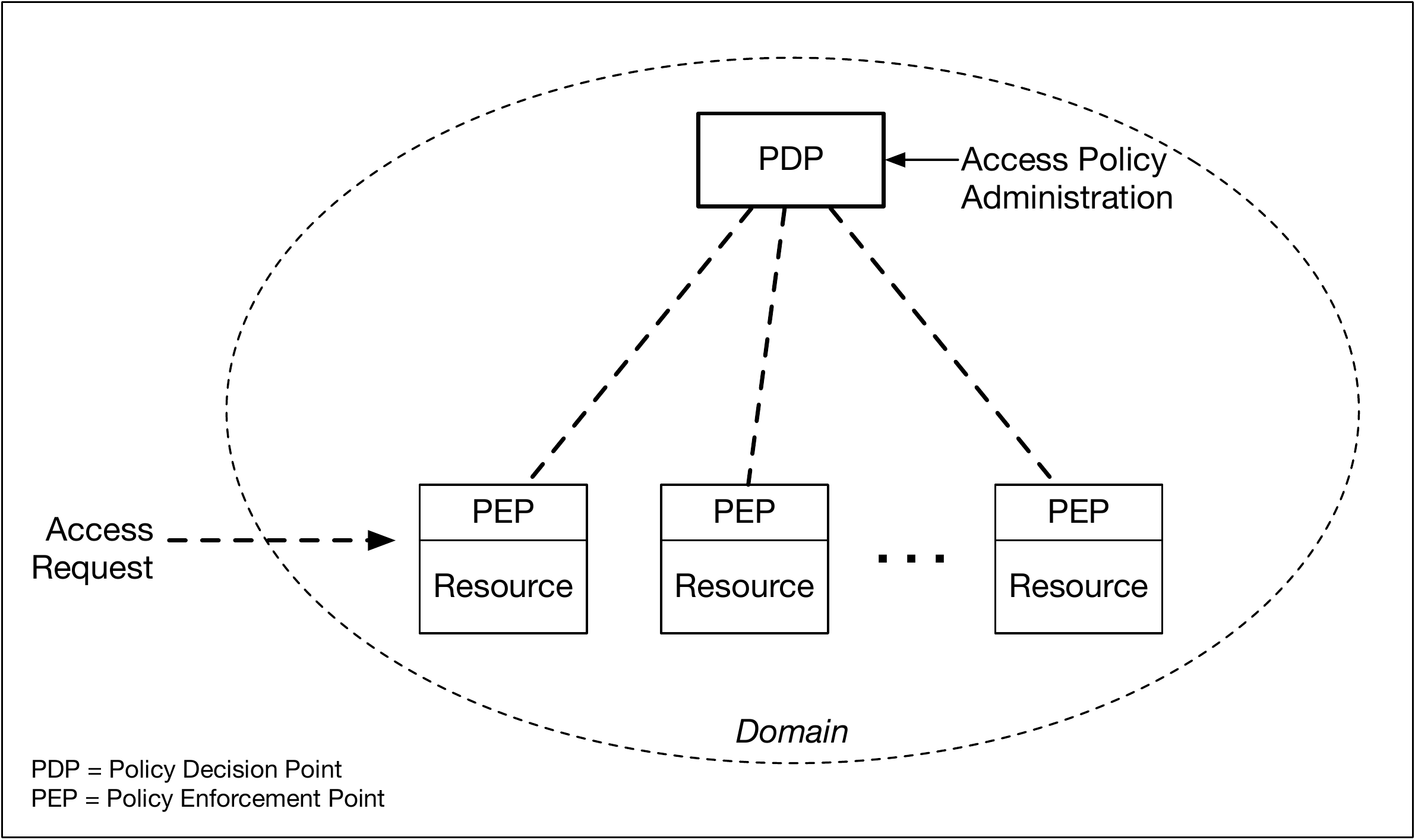}
	%
	%
\caption{Overview of the Policy Based Access Control with PDPs and PEPs}
\label{fig:pdp-pep}
\end{figure}

Access control to resources is also a major concern for enterprises and corporations.
This need became acute with the widespread adoption of Local Area Network (LAN) technology
by enterprise organizations in the 1990s.
The same RBAC model applies also to corporate resources attached to the corporate LAN.
Corporate security policies was therefore expressed in terms
of access-control policies as applied to subjects in certain roles seeking access to objects
residing within a given administrative {\em domain}.
This problem was often referred to as Authentication, Authorization and Audit (AAA)
in the 1990s~\cite{rfc2989}.
Part of the AAA model developed during the 1990s 
was an abstraction of functions pertaining to {\em deciding} access rules,
from functions pertaining to {\em enforcing} them.
Entities which decided on access-rules were denoted as {\em Policy Decision Points} (PDP),
while entities that enforced these access-rules were denoted as {\em Policy Enforcement Points} (PEP)\cite{rfc2753}.
Figure~\ref{fig:pdp-pep} summarizes this abstraction.

The policy-based access control model is foundational to many systems
deployed within enterprises today.
Many solutions, such as Microsoft's Active Directory (AD),
are built on the same model of policy-based access control.
In the case of AD,
a fairly sophisticated cross-domain architecture was developed
which allows an enterprise to logically arrange itself into
dozens to hundreds of interior domains (e.g. each department as a different domain).
Permissions and entitlements for subjects (employees)
are expressed in a comprehensive Privilege Attribute Certificate (PAC)
data structure~\cite{MS-PAC-2018}.

As we will see in the following sections,
although the policy-based access control model works well within the enterprise,
there many challenges today
in carrying-over the model to the ``consumer'' (non-enterprise) use-cases.

\section{Mediated Authentication \& Authorization}
\label{sec:FederatedAuthNAuthz}

Today there are a number of entities on the Internet which
mediate transactions between an individual
(client) and an online service provider.
One key offering of many of these entities
is that {\em mediated authentication} services.

In this section we introduce a slightly modified terminology for clarity of discussion.
This is to avoid employing industry jargon,
which is often inaccurate and a product of historical development.
For example, instead of using
the generic term ``service provider'' for the entities
which offer a broad range of offerings,
we use the more specific terms of {\em Goods \& Services Provider} (GSP)
and {\em Resources Services Provider} (RSP).
The first denotes entities which offer goods (e.g. Amazon),
while the later denote entities that offer computer-related resources
such as cloud-based storage (e.g. DropBox), compute capabilities (e.g. AWS/EC2) and others.
We denote the third party entity seeking to be authenticated or authorized
(prior to accessing a resource)
as the {\em Requesting Party} (RqP).

In order to understand better the notion of federation,
it is useful to review briefly the notion of authentication and authorization
of a client as provided by the third-party providers of these functions (see Figure~\ref{fig:pdp-pep}):
\begin{itemize}

\item	{\em AuthenticatioN Provider} (ANP): A mediated Authentication Provider 
has the task of managing and validating
a user's credential
(e.g. password, keys) on behalf of a GSP entity.
This allows a GSP to be relieved from having to manage
the credentials belonging to users (customers).
Typically the GSP has a business relationship with the ANP,
before the user can perform authentication to the ANP.

There are several variations of the authentication provider protocols.
Generally,
the ANP issues an {\em authentication-token} as proof of the Client's 
successful authentication event at the ANP.
The authentication-token can be delivered to the GSP via the
Client (front channel),
or the token can be delivered directly from the ANP to the GSP (back channel).
An example of these tokens are the Kerberos tickets~\cite{rfc1510} ands
{SAML2.0} login assertions~\cite{SAMLwebsso}.

Today the ANP function is fulfilled by
a category of providers referred to as {\em Identity Providers} (IdP).
The typical consumer-facing IdP issues an identifier (e.g. email address)
and manages the credentials of the user (e.g. change password).
When the user seeks to access services offered by the GSP,
the user is temporarily redirected to the IdP for authentication.
The IdP issues an authentication-token which can then be validated by the GSP.

\item	{\em AuthoriZation Provider} (AZP): A mediated Authorization Provider 
has the task of managing
{\em access policies} (authorizations) pertaining to access to a resource,
such as files, document, media and so on.
The resource typically resides at an RSP entity,
and the owner of the resource sets the access policies at the AZP entity.
A back-channel typically exists between the AZP and the RSPs,
permitting the policy rules and configuration settings
to be communicated from the AZP to the RSPs directly.
Looking at Figure~\ref{fig:pdp-pep}, the AZP implements function of the 
Policy Decision Point (PDP)
while an RSP implements the function of the Policy Enforcement Point (PEP).

When a third party (requesting party) seeks to access a given resource at an RSP,
it must first be authenticated an ANP entity who issues it with an authentication-token.
The ANP is assumed to have a business relationship with the AZP.
The requesting party the wields the authentication-token to the AZP
entity as proof that the requesting party has been authenticated.
The AZP in turn
issues an {\em authorization-token} (e.g. {OAuth2.0} token, Microsoft PAC) 
as a means to convey the access privileges assigned to the user (subject),
for given resource at the RSP.

\end{itemize}

Note that some entities may in fact offer both ANP and AZP functions
within a single platform.
For example, a user that seeks to access an online word-processing platform (e.g. Google Docs)
may be prompted to login and prove their password.
Here the word-processing platform is a resource protected by the AZP function,
while the password validation process is provided by the ANP function. 
As another example,
in the Kerberos system~\cite{rfc1510} the Key Distribution Center (KDC)
acts as both the ANP and AZP
depending on the exchange and message type received from the Client.

\begin{figure}[!t]
\centering
\includegraphics[width=0.9\textwidth, trim={0.0cm 0.0cm 0.0cm 0.0cm}, clip]{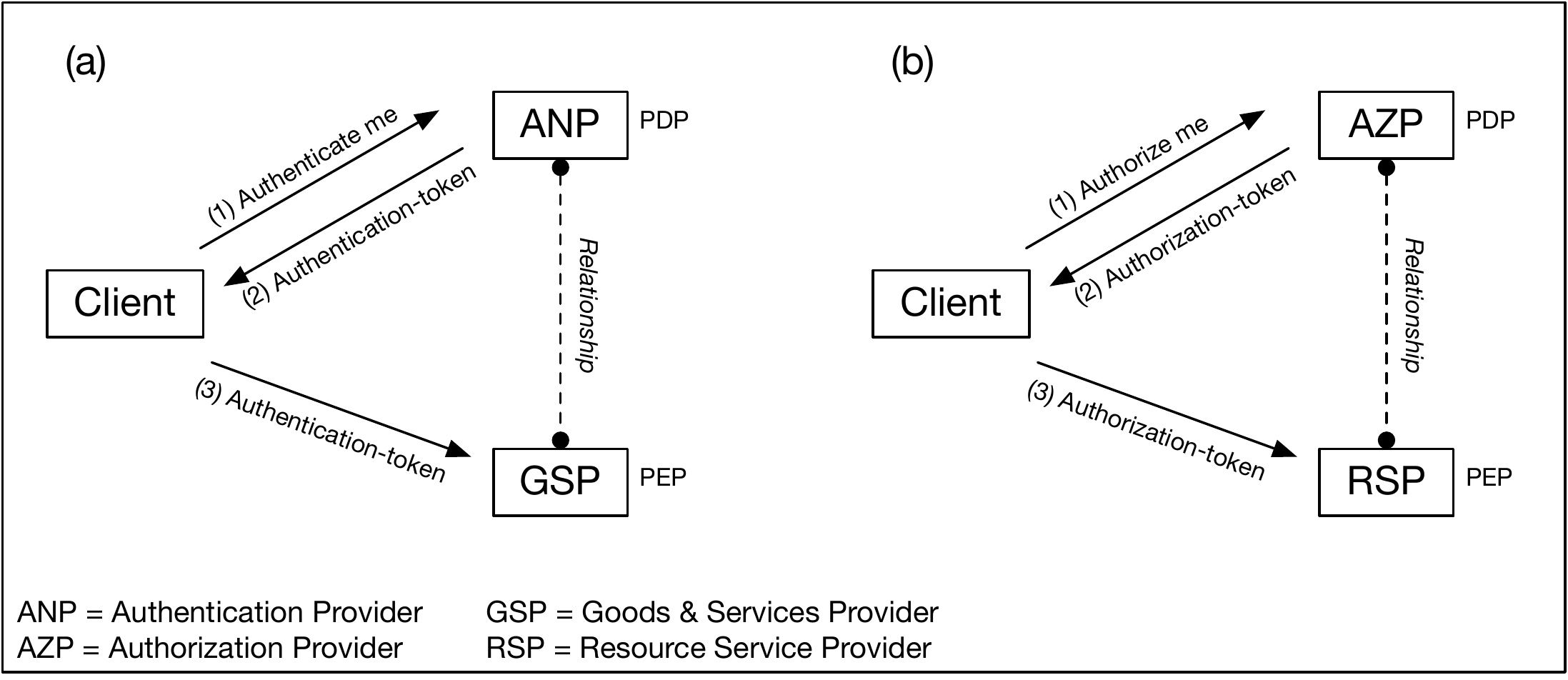}
	%
	%
\caption{Overview of (a) Authentication Provider and (b) Authorization Provider}
\label{fig:authNauthZ}
\end{figure}

In order to scale-up services,
over the years a number of authentication providers (``identity providers'') in the consumer space 
have banded together to form consortiums that provide
its members with a broader reach for their services collectively.
We use the term {\em Authentication Federation} (or identity federation) for this kind of arrangement.
The goal of authentication federation is essentially
to help the GSP entities to ensure that a new or returning user (i.e. customer)
can be authenticated quickly.
To achieve this efficiency, 
a GSP enters into a business relationship either with an ANP who is a member of the federation
or with the federation organization directly.
An authentication federation typically operates under 
a set of bylaws and contracts,
referred to the identity federation {\em Legal Trust Framework} (LTF) for the 
association (e.g. see~\cite{SAFE-BioPharma-2016,OIX2017}).

\section{Federation of Mediated Authorization Services}
\label{sec:DecentFederatedAuthorization}

While the notion of authentication federation 
has evolved over the past decade -- driven primarily
by the large number of online merchants that require one-step
credentials management and payments mechanism
that are consumer-friendly --
mediated authorization services remains a nascent paradigm in the consumer space.
In the enterprise space, mediated authorization is a well known and mature function.
This is notably true in enterprises which employ policy-driven
permissions (entitlements) management for employees, which is usually
tied to Directory Services (e.g. Microsoft Active Directory).

In the consumer space mediated authorization remains nascent
primarily due to the lack of business drivers.
As mentioned above,
we believe that in the near future as private and public institutions
begin to implement the compliance requirements stemming from the various
privacy regulations (e.g. GDPR, CCPA),
these institutions -- and GDPA data controllers -- will have to manage the resources (personal data)
of citizens (data subjects),
and consequently manage access to those resources.

\begin{figure}[!t]
\centering
\includegraphics[width=1.0\textwidth, trim={0.0cm 0.0cm 0.0cm 0.0cm}, clip]{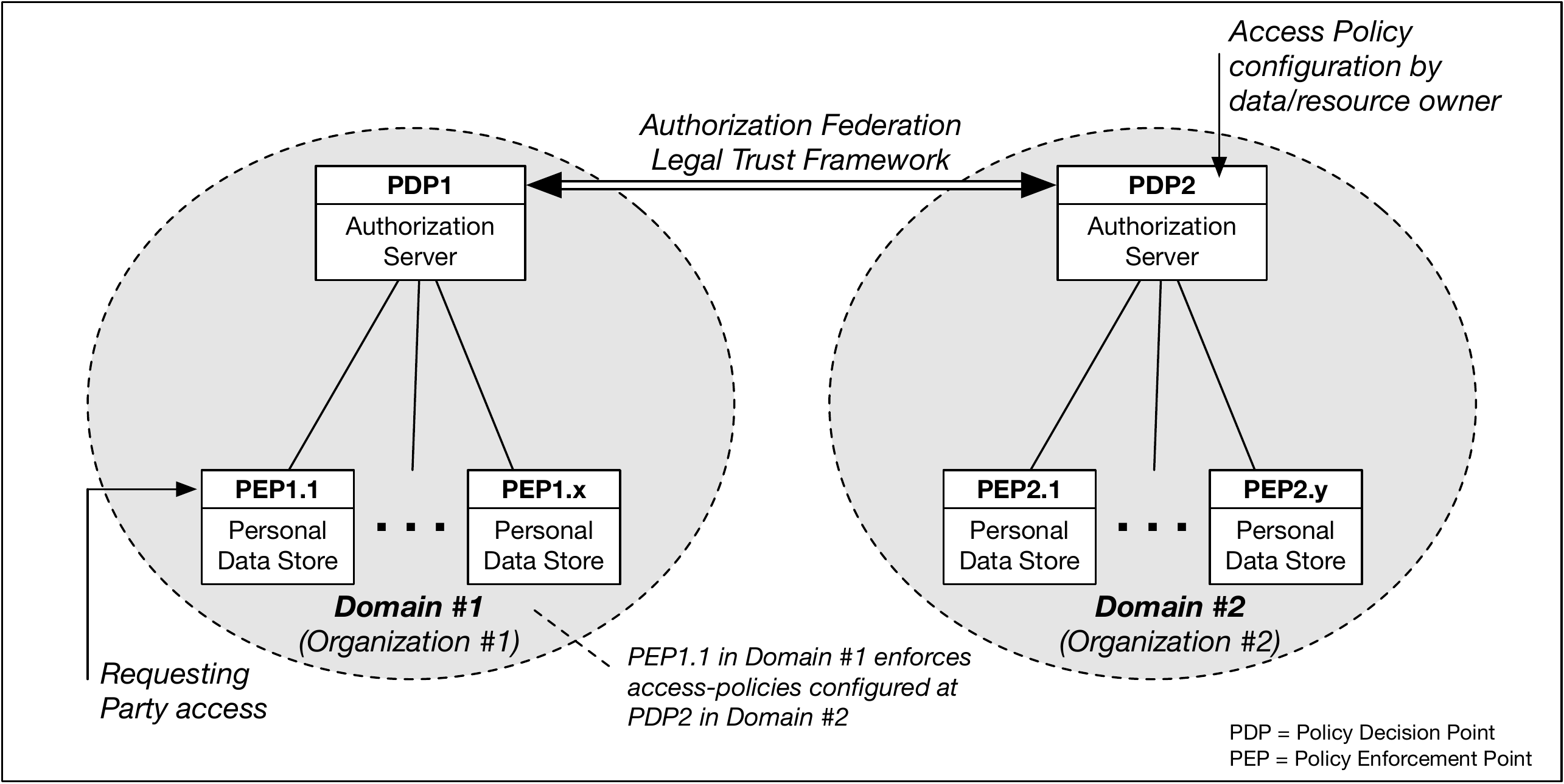}
	%
	%
\caption{Overview of Authorization Federation of Two Domains or Institutions}
\label{fig:decentfedauthz}
\end{figure}

In the following we use the classic 
policy-based resource access control~\cite{FerraioloKuhn1992} 
as our starting point (shown earlier in Figure~\ref{fig:pdp-pep}),
and apply it to a collection of {\em domains},
each representing distinct data controllers (i.e. data-holding institutions).
In Figure~\ref{fig:decentfedauthz},
both Domain~1 and Domain~2 holds resources (personal data) associated with an individual,
which we refer to as the {\em data subject} (or simply {\em subject}) following the GDPR definition~\cite{GDPR}.
The subject has data located at both Domain~1 and Domain~2.
A third party, denoted as the {\em requesting party},
seeks access to the subject's data located in Domain~1.

There are three (3) main goals for a scalable federated authorization model:
\begin{itemize}

\item	{\em Cross-domain policy propagation and enforcement}:
A subject (resource owner) must be able to set access policies
in one domain, and have the policies automatically propagated to 
all domains in the federation that contain the subject's resources
and have those policies enforced locally by each relevant domain.

An example is illustrated in Figure~\ref{fig:decentfedauthz},
where the subject sets access policies at PDP2 in Domain~2
while enforcement occurs also in Domain~1 to resources
in {PEP1.1} where the subject's resources reside.

\item	{\em Decentralization of enforcement}:
Once an access policy is decided at one PDP in one domain,
enforcement within all domains in the federation that contain the subject's resources
must occur automatically without the subject's further involvement.
Each PEP in each relevant domain must operate
independently of other PEPs in the same domain or other domains.

\item	{\em Legal trust framework for cross-domain authorization}:
A legal trust framework must be agreed upon by all domain-owners in the federation, 
one which defines the agreed behavior of PDPs and PEPs
in propagating access policies and enforcing them.

\end{itemize}

An increasingly popular implementation instance of the mediated authorization model
is the user-managed access (UMA) architecture and protocols,
based on the current {OAuth2.0} framework.
This will discussed in the next section in the context of its
application to personal data.

\section{Federated Authorization for Personal Data: UMA}
\label{sec:UMAarchitecture}

The {\em User Managed Access} (UMA) architecture~\cite{UMACORE1.0,UMACORE2.0}  
is an extension of the popular
{\em OAuth2.0} framework~\cite{rfc6749} that provides
basic authorization based on a JSON access token.
The entities and flows in {OAuth2.0} is a subset of the role-based access control model (RBAC)
discussed earlier,
where the three (3) entities are the {\em Client}, the {\em Authorization Server} (AS)
and the {\em Resource Server} (RS).
A client seeking access to a resource (e.g. file) located
at the resource server must first obtain an authorization-token
(called ``access token'') from the authorization server (as the PDP).
Once the token is obtained,
the client must present the token in the access request to the resource server.
As the PEP, the resource server must enforce the access-rule
scoped within the token.

Readers familiar in the classic RBAC model~\cite{FerraioloKuhn1992}
might find the {OAuth2.0} model perplexingly simple and even limiting~\cite{richer2019a}.
However, it is worthwhile to note the motivations as to why {OAuth2.0} emerged and flourished.
One key reason for the success of {OAuth2.0} was the need 
at the time to address the need for applications (apps) in mobile devices 
to obtain renewable authorizations to access low-value accounts (e.g. online calendar, photos, etc)
belonging to the device owner.
Thus, the typical scenario involves the device-owner ``authorizing'' an app software
on the device to access an account (e.g. calendar account, social media account) belonging to the device-owner.
This scenario is often referred to as owner-to-owner (sharing with self) because
the same person owned the device/app and the corresponding online accounts.



\begin{figure}[!t]
\centering
\includegraphics[width=1.0\textwidth, trim={0.0cm 0.0cm 0.0cm 0.0cm}, clip]{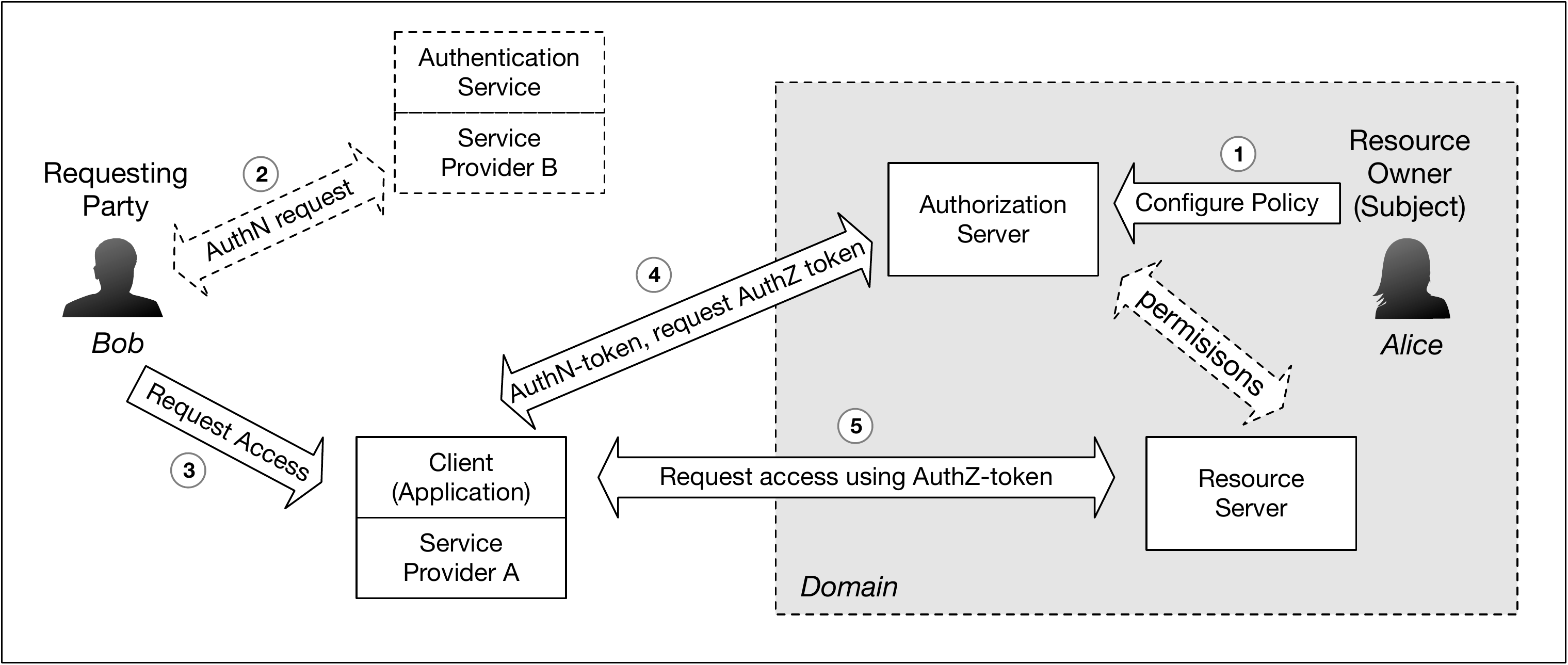}
	%
	%
\caption{Overview of {OAuth2.0} and {UMA}}
\label{fig:uma8party}
\end{figure}

Given the real-world constraints  -- namely the popularity of {OAuth2.0} on social media platforms and mobile devices --
the UMA effort sought to instead re-use the same 3-party {OAuth2.0}  framework
for the positive benefit of individuals.
The UMA philosophy is that individuals should be empowered to control
access to their personal data and resources regardless of where these are located on the Internet.
He or she must be empowered to grant access (issue consent) to a requesting party
and to revoke access (retract consent).
As such, the UMA philosophy has been consistent
with much of the data privacy discourse in the World Economic Forum~\cite{WEF2011,WEF2014}
for the past decade and with the GDPR notion of {\em informed consent}~\cite{GDPR}.

The basic UMA entities are shown in Figure~\ref{fig:uma8party}.
Following {OAuth2.0}, the basic entities are the Client,  Authorization Server (AS)
and the Resource Server (RS).
UMA goes a step further in adding a fourth entity, namely the Requesting Party (RqP),
which can be a person or legal entity.
Additionally, UMA requires that (a) the Requesting Party obtain an authentication-token
from an Identity Provider, and that
(b) both the Requesting Party and the Client (i.e. operator of the Client application)
to obtain separate access-tokens from AS (shown as Step~4 in Figure~\ref{fig:uma8party}).
This last step in effect brings the Client-operator explictly
into the ecosystem and therefore forces it to also obtain consent from the resource owner (the individual).

\begin{figure}[!t]
\centering
\includegraphics[width=1.0\textwidth, trim={0.0cm 0.0cm 0.0cm 0.0cm}, clip]{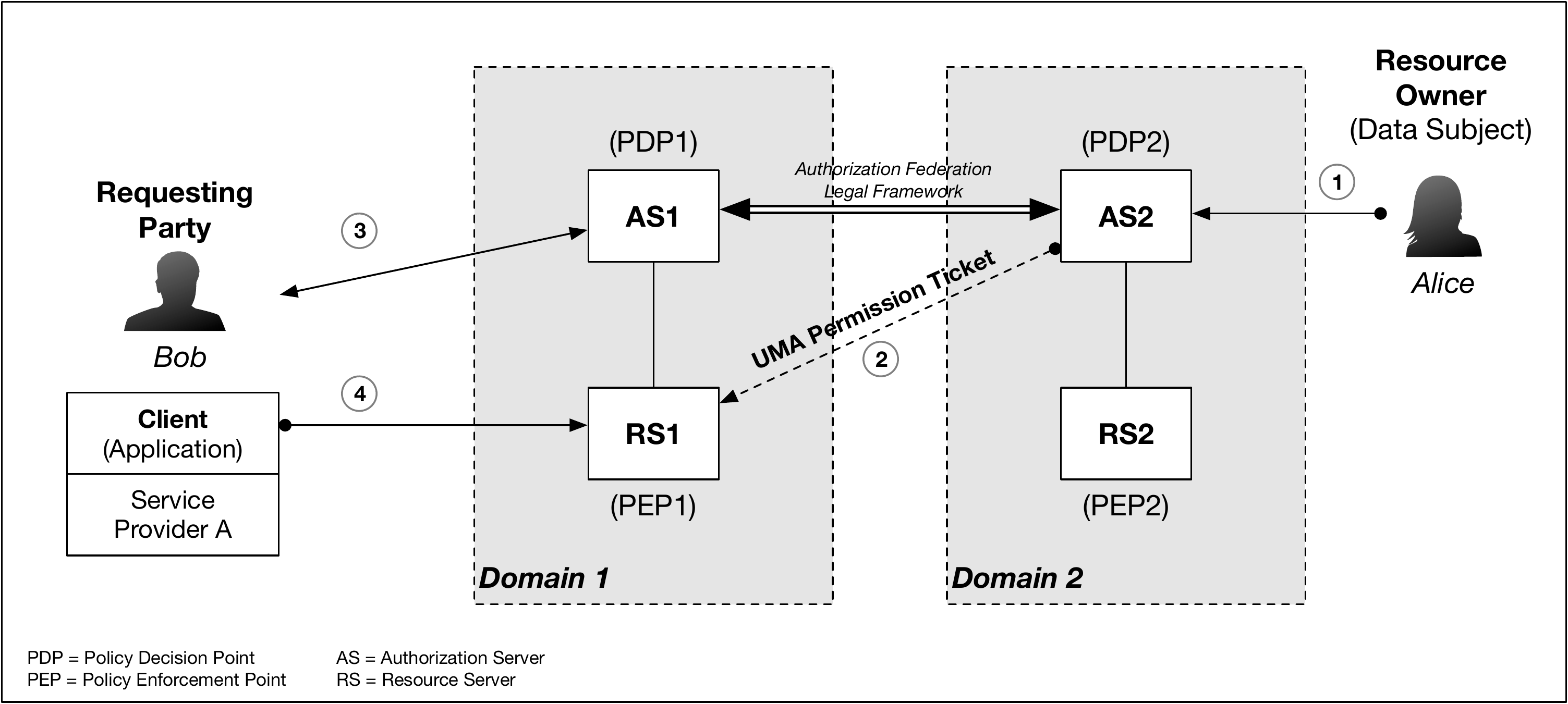}
	%
	%
\caption{Cross-Domain Federated Authorization in {UMA2.0}}
\label{fig:uma-crossdomain}
\end{figure}

Overall, the UMA architecture contributes to advancing the empowerment
of individuals to control access to their resources in the following ways:
\begin{itemize}

\item	{\em Subject-centric policy setting}:
The main focus in UMA is the data-subject (resource owner Alice in Figure~\ref{fig:uma8party})
whose resources (personal data) 
are located at differing domains (institutions) on the Internet.
Access request flows cannot begin until the subject has set access policies on the resources.

\item	{\em Authorization for the client-operator}:
UMA requires both the requesting party (Bob in Figure~\ref{fig:uma8party})
and the client-operator (Service Provider~A in Figure~\ref{fig:uma8party})
to obtain separate access-tokens as proof of authorization to access personal data
belonging to Alice as data-subject (resource owner).
This is because UMA explicitly recognizes that data 
(to which the requesting party has been granted access)
will flow through the IT infrastructure of the client-operator (e.g. cloud infrastructure).
In the current situation (without UMA) the client-operator would be free
to scan and even copy data flowing through 
their systems -- effectively creating a ``data privacy leakage'' in the ecosystem.

\item	{\em Support for cross-domain federated authorization}:
The UMA architecture is designed to support federated authorization 
across multiple domains (see Figure~\ref{fig:uma-crossdomain}).
The standardized data structure used to propagate the access policies
between the Authorization Server (as the PDP) and the multiple
Resource Servers (as the PEPs)
is the {\em permissions ticket}.
UMA itself is agnostic to the policy expression syntax or language 
and can support any policy syntax (e.g. XACML).

Thus, Figure~\ref{fig:uma-crossdomain} illustrates the case
where Alice as the resource owner
has personal data at both RS1 and RS2.
Alice sets her access policies at one location only, namely at AS2.
Bob is seeking access to Alice's data at RS1.
The UMA architecture supports the propagation of 
the permission ticket from the AS2 (PDP2) in the originating-domain 
(Domain~2) to the RS1 (PEP1) in the enforcing-domain (Domain~1).
The signed permission ticket can be propagated from Domain~2 to Domain~1
either directly from AS2 to AS1,
or it can be propagated cross-domain from AS2 to AS1,
followed by a transfer from AS1 to RS1.
This allows RS1 (PEP1) to enforce Alice's access policies
even though Alice may subsequently be off-line.

\end{itemize}

One of the roles of the UMA architecture and protocol specifications
in the context of federating the data controllers
is to provide the technical foundation for these controllers
to establish a legal trust framework for collectively governing
access to personal data at their respective domains.
A group of data controllers in a given vertical or sector (e.g. financial data)
could establish a federation consortium for the purposes 
of providing individuals the capacity
to use data for their own purposes, as stated in~\cite{WEF2014}.
The legal trust framework for the consortium
would specify the various aspect of the membership
roles, duties and obligations with regards to implementing
the UMA-based federated authorization.
For example, this could include
specific the consortium-specific policy expression syntax/language,
name-spaces, data schema, privacy rules, and so on.

\section{Conclusions}
\label{sec:Conclusions}

Today personal data of an individual
is distributed throughout the Internet,
in both private and public institutions,
and increasingly also on the user's devices.
We believe that individuals need meaningful control over their personal data.
Individuals must be empowered by giving them a say in how
data about them is used by organizations and 
by giving individuals the capacity
to use data for their own purposes.

However, in order to provide an individual with practical control over their personal data
the various data controller entities need to collectively
provide federated authorization to govern access to that personal data.
The overall goal of federated authorization is to empower the individual
to set access policies (consent permission) at one location (e.g. at one data controller)
and have the access policies propagated to other data controllers
and be enforced there also.
In this way, the individual is relieved from having to log-in to
many sites for the purpose or configuring the access policies multiple times.

The User Managed Access (UMA) architecture implements such a federated authorization model.
UMA provides subject-centric policy setting capabilities that allows the individual
to configure access policies (consent) and to retract those policies.
It is designed to support federated authorization 
across multiple domains,
which in this case corresponds to the various organizations and institutions which hold personal data.

\section*{Acknowledgments}

We thank Eve Maler for leadership and vision
in the UMA Working Group (UMA~WG) within the Kantara Initiative since its inception in 2009.
We also thank the various UMAnitarians that have
worked collaboratively since the start of the UMA~WG:
Maciej Machulak,
Domenico Catalano,
George Fletcher,
Mike Schwartz,
Justin Richer,
Mark Lizar,
Tim Reiniger
and Colin Wallis.
Finally, we thank the following at MIT: 
Sandy Pentland, Alex Lipton, Cameron Kerry, Justin Anderson, Denis Babani and Stephen Buckley.



\end{document}